\documentclass[twocolumn,showpacs]{revtex4}

\usepackage{graphicx}
\usepackage{dcolumn}
\usepackage{bm}
\usepackage{hyperref}

\begin{document}

\title{Initial data for binary neutron stars with arbitrary spins}

\author{Wolfgang Tichy}
\affiliation{Department of Physics, Florida Atlantic University,
             Boca Raton, FL  33431, USA}




%
\newcommand\be{\begin{equation}}
\newcommand\ba{\begin{eqnarray}}

\newcommand\ee{\end{equation}}
\newcommand\ea{\end{eqnarray}}
\newcommand\p{{\partial}}
\newcommand\remove{{{\bf{THIS FIG. OR EQS. COULD BE REMOVED}}}}
%

\begin{abstract}

In general neutron stars in binaries are spinning.
Due to the existence of millisecond
pulsars we know that these spins can be substantial.  We argue that spins
with periods on the order a few dozen milliseconds could influence the late
inspiral and merger dynamics.  Thus numerical simulations of the last few
orbits and the merger should start from initial conditions that allow for
arbitrary spins.  We discuss quasi-equilibrium approximations one can make in
the construction of binary neutron star initial data with spins.  Using
these approximations we are able to derive two new matter equations.  As in
the case of irrotational neutron star binaries one of these equations is
algebraic and the other elliptic.  If these new matter equations are solved
together with the equations for the metric variables following the
Wilson-Mathews or conformal thin sandwich approach one can construct neutron
star initial data.  The spin of each star is described by a rotational
velocity that can be chosen freely so that one can create stars in arbitrary
rotation states.  Our new matter equations reduce to the well known limits
of both corotating and irrotational neutron star binaries.

\end{abstract}

\maketitle

\section{Introduction}

Several gravitational wave detectors such as
LIGO~\cite{LIGO:2007kva,LIGO_web},
Virgo~\cite{VIRGO_FAcernese_etal2008,VIRGO_web}
or GEO~\cite{GEO_web} have been operating over the last few years,   
while several others are in the
planning or construction phase~\cite{Schutz99}. One of the most promising
sources for these detectors are the inspirals and mergers of binary neutron
stars. In order to make predictions about the last few orbits
and the merger of such systems,
fully non-linear numerical simulations of the Einstein
Equations are required. To start such simulations we need initial
data that describe the binary a few orbits before merger.
The emission of gravitational waves
tends to circularize the orbits~\cite{Peters:1963ux,Peters:1964}.
Thus, during the inspiral, we expect the two neutron stars to be
in quasi-circular orbits around each other with a radius
that shrinks on a timescale much larger than the orbital
timescale. This means that the initial data should
have an approximate helical Killing vector $\xi^{\mu}$.
To incorporate these ideas we will use the 
Wilson-Mathews approach~\cite{Wilson95,Wilson:1996ty},
which is also known as conformal thin sandwich formalism~\cite{York99},
for the metric variables.
The Wilson-Mathews approach has already been successfully used by
several groups together with matter equations 
describing the neutron stars in either 
corotating~\cite{Baumgarte:1997xi,Baumgarte:1997eg,
Mathews:1997pf,Marronetti:1998xv,Tichy:2009yr}
or irrotational~\cite{Bonazzola:1998yq,Gourgoulhon:2000nn,Marronetti:1999ya,
Uryu:1999uu,Marronetti:2003gk,Taniguchi:2002ns,Taniguchi:2003hx,
Uryu:2005vv,Uryu:2009ye}
states. There have also been attempts to include intermediate rotation
states~\cite{Marronetti:2003gk,Baumgarte:2009fw}. However, as we will
discuss in more detail later, these approaches have certain drawbacks,
because they do not correctly solve the Euler equation for the fluid.
Thus, so far there is no canonical formalism to describe neutron star 
binaries with arbitrary spins.
As pointed out by Bildsten and Cutler~\cite{Bildsten92},
the two neutron stars cannot be tidal locked,
because the viscosity of neutron star matter is too low.
Hence barring other effects like magnetic dipole radiation the spin of
each star remains approximately constant.
This means that initial data sequences of corotating configurations
for different separations cannot be used to approximate the inspiral of two
neutron stars. 
On the other hand, sequences of irrotational configurations can be used
to approximate the inspiral of two neutron stars without spin.
This fact explains why irrotational initial data are far more popular
today. Nevertheless, astrophysical neutron stars will have a non-zero spin.
Therefore a corotating configuration at some particular separation
does have its place as a possible initial configuration with spin. 
It will just not remain corotating during the subsequent time evolution.
Of course real neutron stars 
will likely have spins that have periods different
from the orbital period, and the spin direction may not be aligned with the
orbital angular momentum. Thus it would be highly desirable to have
a formalism that can be used to generate initial data for arbitrary initial
spins. 

In order to judge how important spins might be let us discuss a few order of
magnitude estimates.
A typical neutron star has a mass of about 1.4 solar masses ($M_{\odot}$) 
and a radius on the order of 15km.
From Kepler's law the orbital period
\begin{equation}
P_o \sim  \left(\frac{d}{50\mbox{km}}\right)^{3/2} 
          \left(\frac{M_{\odot}}{M}\right)^{1/2}  6\mbox{ms} 
\end{equation}
is on the order of a few milliseconds during the last orbit before
merger where the separation $d\sim 50\mbox{km}$.
Thus systems with spin periods that are much larger than $P_o$
should be treatable as approximately irrotational, while systems
with spin periods of a few milliseconds 
(such as millisecond pulsars) cannot be regarded as irrotational.
Another way of judging how important spins could be during the evolution
is to look at the dimensionless spin magnitude.
If we assume that the spin $S$ of a neutron star 
with mass $m$ and radius $R$ is related to its spin period $P$ by
$S= I (2\pi/P)$ 
with $I\sim m R^2$ 
we find that the dimensionless spin has a magnitude of
\begin{equation}
\frac{S}{m^2}
\sim \left(\frac{R}{15\mbox{km}}\right)^2 
                    \frac{M_{\odot}}{m} \frac{3 \mbox{ms}}{P} .
\end{equation}
Thus millisecond pulsars have a dimensionless spin of order one.
As in the case of binary black holes~\cite{Campanelli:2006uy,
Campanelli:2006fy,Campanelli:2007ew,
Tichy:2007hk,Marronetti:2007wz,Tichy:2008du}, 
spins of this magnitudes could have a
significant influence on the merger dynamics. This means that
neutron stars with spin periods of a few dozen milliseconds
or less should not be considered irrotational. One could of course
imagine that the neutron stars spin down before they enter
the strongly relativistic regime of the last few orbits before
merger that is usually considered in numerical relativity simulations.
In order to address this question let us look at the famous double pulsar
PSR J0737-3039 which is the only neutron star binary where
both spin periods and spin down rates are known~\cite{Lyne:2004cj}. 
Star A has mass $m_A=1.34 M_{\odot}$ and spin period $P_A=23\mbox{ms}$,
while star B has $m_B=1.25 M_{\odot}$ and $P_B=2.8\mbox{s}$. The orbital
period is $P_o=2.4\mbox{h}$. 
From these numbers one derives that the system will merge in about
$85\mbox{My}$ due to the emission of gravitational waves. Both stars are
currently spinning down at a rate of
$\dot{P}_A=1.7\times 10^{-18}$ and 
$\dot{P}_B=8.8\times 10^{-16}$~\cite{Lyne:2004cj}.
If one assumes that this spin down is due to magnetic dipole radiation
and defines the characteristic ages given by 
$\tau_A=P_{A}/(2\dot{P}_{A})= 210\mbox{My}$
and $\tau_B=P_{B}/(2\dot{P}_{B})= 50\mbox{My}$,
one finds that the spin period of each star obeys~\cite{Shapiro83}
\begin{equation}
P_{A/B}(t) = P_{A/B}(0) \sqrt{1+\frac{t}{\tau_{A/B}}} ,
\end{equation}
where the time $t=0$ is the time today. From this it is clear that the periods
at merger (at $t=85\mbox{My}$) will be 
$P_{A}(t)=27\mbox{ms}$ and $P_{B}(t)=4.6\mbox{s}$. 
Thus star A will not spin down enough to be well approximated by an 
irrotational configuration by this time. This example shows 
that neutron stars in binary systems can have appreciable spins
a few orbits before merger.
Of course since only about ten binary neutron stars have
been observed so far~\cite{Lorimer:2005bw} it is not clear yet
how common binary neutron stars with high spins are. However, since there 
are numerous millisecond pulsars it seems reasonable to expect that
neutron stars in binaries can also have millisecond spin periods.
Hence the widely held belief that only irrotational configurations are
realistic is not necessarily correct. It is thus necessary to develop
initial data for binary neutron stars with arbitrary spins.
In the next sections we will describe what approximation one can make
to derive a formalism that allows for this possibility. We will see
that our new equations reduce to well known accepted results in both
the corotating and irrotational cases.

Throughout we will use units where $G=c=1$.
Latin indices such as $i$ run from 1 to 3 and denote spatial indices,
while Greek indices such as $\mu$ run from 0 to 3 and denote spacetime
indices.
The paper is organized as follows. 
Sec.~\ref{BNSequations} lists the General Relativistic equations
that govern binary neutron stars described by perfect fluids.
We use three approximate quasi-equilibrium conditions to simplify these
equations. We find two new matter equations that allow as to
set up binary neutron stars with arbitrary spins.
In Sec.~\ref{NewtonLimit} we consider the Newtonian limit of our new
equations.
We conclude with a discussion of our method in Sec.~\ref{discussion}.
In the appendix we discuss our quasi-equilibrium conditions 
for a simple case.

\section{Binary neutron stars with arbitrary rotation states}
\label{BNSequations}

In this section we describe the equations governing
binary neutron stars in arbitrary rotation states in General Relativity.
The equations for the metric and matter variables discussed in 
subsections \ref{ADMdecomposition}, 
\ref{matter_eqs}, \ref{gK_Decomposition} and \ref{Quasiequil_geometry}
are well known. Our new results concerning quasi-equilibrium
conditions for neutron stars with arbitrary rotation states
are presented in subsections \ref{Quasiequil_matter}
and \ref{simps_and_BCs}.

\subsection{ADM decomposition of Einstein's equations}
\label{ADMdecomposition}

We use the Arnowitt-Deser-Misner (ADM) decomposition of Einstein's
equations (see e.g.~\cite{Misner73}) and introduce the 3-metric
\begin{equation}
\gamma_{\mu\nu} = g_{\mu\nu} + n_{\mu} n_{\nu} .
\end{equation}
Here $g_{\mu\nu}$ is the spacetime metric and $n_{\mu}$ is the unit normal
to the $t=const$ hypersurface. The line element is then
\begin{equation}
ds^2 = -\alpha^2 dt^2 + \gamma_{ij} (dx^i + \beta^i dt)(dx^j + \beta^j dt) ,
\end{equation}
where the lapse $\alpha$ and shift $\beta^i$ are related to $n_{\mu}$
via
\begin{equation}
n^{\mu}=\left(1/\alpha , -\beta^i/\alpha \right)  \ \ \ \         
n_{\mu}=\left(-\alpha , 0,0,0 \right) 
\end{equation}
The extrinsic curvature is defined by
\begin{equation}
K_{ij} = -\frac{1}{2\alpha}
         (\partial_t \gamma_{ij} - \pounds_{\beta} \gamma_{ij}),
\end{equation}
With these definitions Einstein's equations split into
the evolution equations
\begin{eqnarray}
\label{evo0}
\partial_t \gamma_{ij} &=& -2\alpha K_{ij} + \pounds_{\beta} \gamma_{ij} \nonumber \\
\partial_t K_{ij} &=& \alpha (R_{ij} - 2 K_{il} K^l_j + K K_{ij})
 - D_i  D_j \alpha + \pounds_{\beta} K_{ij} \nonumber \\
 &&- 8\pi S_{ij} + 4\pi\gamma_{ij}(S-\rho)
\end{eqnarray}
and the Hamiltonian and momentum constraint equations
\begin{eqnarray}
\label{ham0}
R -  K_{ij}  K^{ij} + K^2   &=& 16\pi\rho \nonumber \\
\label{mom0}
D_j(K^{ij} - \gamma^{ij} K) &=& 8\pi j^i .
\end{eqnarray}
Here $R_{ij}$ and $R$ are the Ricci tensor and scalar computed from
$\gamma_{ij}$, $D_i$ is the derivative operator compatible 
with $\gamma_{ij}$ and all indices here are raised and lowered
with the 3-metric $\gamma_{ij}$.
The source terms $\rho$, $j^i$, $S_{ij}$ and 
$S=\gamma^{ij}S_{ij}$ are projections of the stress-energy 
tensor $T_{\mu\nu}$ given by
\begin{eqnarray}
\label{mattervars}
\rho   &=& T_{\mu\nu} n^{\mu} n^{\nu} \nonumber \\
j^i    &=& -T_{\mu\nu} n^{\mu} \gamma^{\nu i} \nonumber \\
S^{ij} &=& T_{\mu\nu} \gamma^{\mu i} \gamma^{\nu j}
\end{eqnarray}
and correspond to the energy density, flux and stress-tensor.

\subsection{Matter equations}
\label{matter_eqs}

We assume that the matter in both stars is a perfect fluid with a
stress-energy tensor
\begin{equation}
T^{\mu\nu} = [\rho_0(1+\epsilon) + P] u^{\mu} u^{\nu} + P g^{\mu\nu}.
\end{equation}
Here $\rho_0$ is the mass density (which is proportional the number
density of baryons), $P$ is the pressure, $\epsilon$ is the internal energy
density divided by $\rho_0$ and $u^{\mu}$ is the 4-velocity of the fluid.
The matter variables in Eq.(\ref{mattervars}) are then
\begin{eqnarray}
\label{fluid_matter}
\rho   &=& \alpha^2 [\rho_0(1+\epsilon) + P] u^0 u^0 - P \nonumber \\
j^i    &=& \alpha[\rho_0(1+\epsilon) + P] u^0 u^0
           (u^i/u^0 + \beta^i) \nonumber \\
S^{ij} &=& [\rho_0(1+\epsilon) + P]u^0 u^0 
           (u^i/u^0 + \beta^i) (u^j/u^0 + \beta^j) \nonumber \\
       & &  + P \gamma^{ij}
\end{eqnarray}

From $\nabla_{\nu} T^{\mu\nu} =0 $ we obtain the relativistic
Euler equation
\begin{equation}
\label{Euler0}
[\rho_0(1+\epsilon) + P] u^{\nu} \nabla_{\nu} u^{\mu} 
= -(g^{\mu\nu} + u^{\mu} u^{\nu}) \nabla_{\nu} P, 
\end{equation}
which together with the continuity equation
\begin{equation}
\label{continuity0}
\nabla_{\nu} (\rho_0 u^{\nu}) = 0
\end{equation}
governs the fluid. 

In order simplify the problem we assume that internal energy $\epsilon$
is a function of $\rho_0$ alone (which implies a temperature of zero),
and use a polytropic equation of state 
\begin{equation}
\label{polytrop}
P = \kappa \rho_0^{1+1/n} .
\end{equation}
We also introduce the specific enthalpy
\begin{equation}
\label{specenthalpy}
h = 1 + \epsilon + P/\rho_0 .
\end{equation}
Changes in $h$ at zero temperature obey
\begin{equation}
\label{dh}
dh = dP/\rho_0 .
\end{equation}
Using Eqs.~(\ref{specenthalpy}) and (\ref{dh}) we can rewrite 
the Euler Eq.~(\ref{Euler0}) as
\begin{equation}
\label{Euler1}
u^{\mu} \nabla_{\mu} \tilde{u}_{\nu} + \nabla_{\nu} h = 0 ,
\end{equation}
where 
\begin{equation}
\tilde{u}^{\nu} = h u^{\nu} .
\end{equation}

It is often convenient to introduce the dimensionless ratio
\begin{equation}
q = P/\rho_0 ,
\end{equation}
which we can use to write
\begin{eqnarray}
\label{hrhoPS_q}
h        &=& (n+1) q + 1 \nonumber \\
\rho_0   &=& \kappa^{-n} q^n \nonumber \\
P        &=& \kappa^{-n} q^{n+1} \nonumber \\
\epsilon &=& n q .
\end{eqnarray}

\subsection{Decomposition of 3-metric and extrinsic curvature}
\label{gK_Decomposition}

As in~\cite{Wilson95,Wilson:1996ty}
the 3-metric $\gamma_{ij}$ is decomposed into a
conformal factor $\psi$ and a conformal metric
$\bar{\gamma}_{ij}$ such that
\begin{equation}
\gamma_{ij} = \psi^4 \bar{\gamma}_{ij} .
\end{equation}
The extrinsic curvature is split into its trace $K$ and its
tracefree part $A_{ij}$ by writing it as
\begin{equation}
K_{ij} = A_{ij} + \frac{1}{3} \gamma_{ij} K
\end{equation}

\subsection{Quasi-equilibrium assumptions for the metric variables}
\label{Quasiequil_geometry}

We now make some additional simplifying assumptions.
First we assume that the binary is in an approximately circular
orbit and that the spins of each star remain approximately constant.
As in the case of binary black holes (see
e.g.~\cite{Tichy:2003zg,Tichy:2003qi})
this implies the existence of an approximate helical Killing vector
$\xi^{\mu}$ with $\pounds_{\xi} g_{\mu\nu} \approx 0$.
In order to clarify the meaning of the approximate sign we now briefly
discuss two cases.

If both spins are parallel to the orbital angular momentum
we have $\pounds_{\xi} g_{\mu\nu} = O(P_o/T_{ins})$, 
where we assume the inspiral timescale $T_{ins}$ to be much 
longer than the orbital timescale $P_o$. I.e. in a corotating coordinate
system all metric time derivatives are of order $O(P_o/T_{ins})$
and thus small.
For arbitrary spins the situation becomes more complicated. We can again
use corotating coordinates, but in this coordinate system the spin vectors
will be precessing on an orbital timescale $P_o$. This means there are
matter currents that change on a timescale $P_o$, while the matter
distribution itself only changes on the inspiral timescale $T_{ins}$.
In this case it is useful to consider gravity to be made up of
gravitoelectric and gravitomagnetic fields~\cite{Braginsky:1976rb,Misner73}.
The gravitoelectric parts of the metric are sourced by the matter
distribution and thus change only on the timescale $T_{ins}$, while
the gravitomagnetic parts of the metric are sourced by matter currents
and thus change on the shorter timescale $P_o$.
However, the gravitomagnetic parts are smaller 
than the gravitoelectric parts by $O(v/c)$~\cite{Braginsky:1976rb,Misner73}.
Thus we now have $\pounds_{\xi} g_{\mu\nu} = O(v/c) \approx 0$,
where we assume that the orbital velocity $v$ is smaller than the speed of
light.

An approximate helical Killing vector with 
$\pounds_{\xi} g_{\mu\nu} \approx 0$ implies that
\begin{equation}
\label{geom_equil}
\pounds_{\xi} \bar{\gamma}_{ij} \approx \pounds_{\xi} K \approx 0 ,
\end{equation}
which is what we will need to assume here for the metric variables.
In a corotating coordinate system where the time evolution vector
lies along $\xi^{\mu}$, the time derivatives of these metric variables
are then equal to zero. 
From $\partial_t \bar{\gamma}_{ij} = 0$ it follows that
\begin{equation}
A^{ij} = \frac{1}{2\psi^4 \alpha}(\bar{L}\beta)^{ij} , 
\end{equation}
where
\begin{equation}
(\bar{L}\beta)^{ij} = 
\bar{D}^i \beta^j + \bar{D}^j \beta^i -\frac{2}{3}  \bar{D}_k \beta^k ,
\end{equation}
and $\bar{D}_k$ is the derivative operator compatible 
with $\bar{\gamma}_{ij}$.
The assumption $\partial_t K = 0$ together with the evolution equation
of $K$ (derived from Eq.~(\ref{evo0})) implies
\begin{eqnarray}
\label{dK0}
\psi^{-5}[\bar{D}_k\bar{D}^k (\alpha\psi) - \alpha \bar{D}_k\bar{D}^k \psi]
&=& \alpha (R + K)^2 + \beta^i \bar{D}_i K \nonumber \\
& &  +4\pi\alpha(S-3\rho) .
\end{eqnarray}

\subsection{Quasi-equilibrium assumptions for the matter variables}
\label{Quasiequil_matter}

In an inertial frame (i.e. a frame with $\lim_{r\to\infty}\beta^i = 0$)
the approximate helical 
Killing vector has the components
\begin{equation}
\xi^{\mu} = 
\left( 1, -\Omega [x^2 - x^2_{CM}], \Omega [x^1 - x^1_{CM}], 0  \right) .
\end{equation}
Here $x_{CM}^i$ denotes the center of mass position of the system
(which can be obtained from surface integrals at infinity e.g.
Eq.~(20.11) in~\cite{Misner73}), and $\Omega$
is the orbital angular velocity, which we have chosen to lie along
the $x^3$-direction.
Following Shibata~\cite{Shibata98} we decompose the fluid velocity $u^{\mu}$
into a piece along $\xi^{\mu}$ and a spatial vector $V^{\mu}$
and write
\begin{equation}
\label{V-def}
u^{\mu} = u^0 \left( \xi^{\mu} + V^{\mu} \right) ,
\end{equation}
where $u^0 = -u^{\mu} n_{\mu}/\alpha$.

In terms of $\xi^{\mu}$ and $V^{\mu}$
the fluid equations (\ref{continuity0}) and (\ref{Euler1}) can be recast as
\begin{equation}
\label{continuity2}
D_i \left( \rho_0 \alpha u^0 V^i \right) + 
\alpha\left[ \pounds_{\xi}(\rho_0 u^0) + 
             \rho_0 u^0 g^{\mu\nu} \pounds_{\xi} g_{\mu\nu}  \right] = 0 
\end{equation}
and
\begin{equation}
\label{Euler2}
D_i \left( \frac{h}{u^0} + ^{(3)}\!\tilde{u}_k  V^k \right) +
V^k \left( D_k ^{(3)}\!\tilde{u}_i - D_i ^{(3)}\!\tilde{u}_k \right) +
\gamma_i^{\nu} \pounds_{\xi} \tilde{u}_{\nu} = 0 , 
\end{equation}
where
\begin{equation}
^{(3)}\!\tilde{u}^i = \gamma^i_{\nu} \tilde{u}^{\nu} .
\end{equation}

When one constructs neutron star initial data for corotating or irrotational
configurations one usually assumes that the Lie derivatives
of all matter variables with respect to $\xi^{\mu}$ 
vanish~\cite{Baumgarte:1997eg,Shibata98,Teukolsky98}. However, for arbitrary
spins this may not be the best approximation, since the
portion of the fluid velocity responsible for the star's spin is not 
constant along $\xi^{\mu}$ if the spin remains constant
while the stars orbit around each other. So we should not assume that
$\pounds_{\xi} \tilde{u}^{\mu}$ vanishes. 
Rather we will split $\tilde{u}^{\mu}$ into an
irrotational and a rotational part and assume that only the Lie derivative
of the irrotational part vanishes.
In the irrotational (zero spin) case we have 
$D_i ^{(3)}\!\tilde{u}_j - D_j ^{(3)}\!\tilde{u}_i = 0$ and thus
$^{(3)}\!\tilde{u}_i$ is derivable from a potential.
For general rotation states we write 
\begin{equation}
\label{utilde-split}
^{(3)}\!\tilde{u}^i = D^i \phi + w^i ,
\end{equation}
so that $D^i \phi$ and $w^i$ denote the irrotational and rotational
pieces of the velocity. In order to assure that $w^i$ is purely 
rotational one usually requires that
\begin{equation}
\label{w_pure_rot}
D_i w^i = 0 .
\end{equation}
In subsection \ref{simps_and_BCs} we will show how one can choose
$w^i$ such that Eq.~(\ref{w_pure_rot}) is satisfied. 
However, Eq.~(\ref{w_pure_rot}) is not explicitly used in
any of the derivations in this subsection.

Note that once $^{(3)}\!\tilde{u}_i$ is known 
$\tilde{u}^0 = -\tilde{u}^{\mu} n_{\mu}/\alpha$ can be obtained from
$\tilde{u}^{\mu} \tilde{u}_{\mu} = -h^2$. If we 
choose $w^{\mu} n_{\mu} = 0$ the split of $^{(3)}\!\tilde{u}_i$
in Eq.~(\ref{utilde-split}) can be extended to
\begin{equation}
\tilde{u}^{\mu} = \nabla^{\mu} \phi + w^{\mu} ,
\end{equation}
where the time dependence of $\phi$ is now chosen such that it
satisfies $\nabla^{0} \phi = \tilde{u}^0$

In order to simplify Eqs.~(\ref{continuity2}) and (\ref{Euler2})
we now assume that 
\begin{equation}
\pounds_{\xi}(\rho_0 u^0) \approx \pounds_{\xi} g_{\mu\nu} \approx 0
\end{equation}
but we will not assume that $\pounds_{\xi} \tilde{u}_{\nu}$ vanishes as
well. Instead we assume that
\begin{equation}
\label{assumption1}
\gamma_i^{\nu} \pounds_{\xi} \left(\nabla_{\nu}\phi\right) \approx 0 ,
\end{equation}
so that the time derivative of the irrotational piece of the
fluid velocity vanishes in corotating coordinates.
Furthermore we also assume that
\begin{equation}
\label{assumption2}
\gamma_i^{\nu} \pounds_{\bar{\xi}} w_{\nu} \approx 0 ,
\end{equation}
where we have defined
\begin{equation}
\label{xibar-def}
{\bar{\xi}}^{\mu} = \frac{\nabla^{\mu} \phi}{\tilde{u}^0} .
\end{equation}
The assumption in Eq.~(\ref{assumption2}) describes the fact that the
rotational piece of the fluid velocity (which gives rise to the spin)
is constant along $\bar{\xi}^{\mu}$ which is parallel to the
worldline of the star center.
Defining
\begin{equation}   
\label{Deltaxi-def}
\Delta\xi^{\mu}=\xi^{\mu} - \bar{\xi}^{\mu} = (0, \Delta k^i) .
\end{equation}
and using Eqs.~(\ref{assumption1}) and (\ref{assumption2}) 
the Lie derivative term in Eq.~(\ref{Euler2}) can be written as
\begin{eqnarray}
\label{fi-term}
\gamma_i^{\nu} \pounds_{\xi} \tilde{u}_{\nu}
&\approx& \gamma_i^{\nu} \pounds_{\xi} w_{\nu}
 =        \gamma_i^{\nu} \pounds_{\bar{\xi}+\Delta\xi} w_{\nu} \nonumber \\
&\approx& \gamma_i^{\nu} \pounds_{\Delta\xi} w_{\nu}
 =        ^{(3)}\!\pounds_{\Delta k} w_i .
\end{eqnarray}
Here $^{(3)}\!\pounds$ is the Lie derivative in 3 dimensions.
Thus Eqs.~(\ref{continuity2}) and (\ref{Euler2}) simplify and 
can be rewritten as
\begin{equation}
\label{continuity3}
D_i \left( \rho_0 \alpha u^0 V^i \right) = 0
\end{equation}
and
\begin{equation}
\label{Euler3}
D_i \left( \frac{h}{u^0} + V^k D_k \phi \right) +
^{(3)}\!\pounds_{V + \Delta k} w_i 
= 0 . 
\end{equation}
In order to further simplify Eq.~(\ref{Euler3}) note that 
\begin{equation}   
V^i + \Delta k^i = \frac{u^i}{u^0} - \xi^i + \Delta \xi^i
= \frac{\tilde{u}^i}{\tilde{u}^0} - \bar{\xi}^{i} 
= \frac{w^i}{\tilde{u}^0} ,
\end{equation}
which follows from Eqs.~(\ref{V-def}), (\ref{utilde-split}),
(\ref{xibar-def}) and (\ref{Deltaxi-def}). 
Hence
\begin{equation}
\label{assumption3}
^{(3)}\!\pounds_{V + \Delta k} w_i 
= 
\frac{w_i}{\tilde{u}^0} \ 
^{(3)}\!\pounds_{\frac{w}{\tilde{u}^0}}\tilde{u}^0 +
w^k \ ^{(3)}\!\pounds_{\frac{w}{\tilde{u}^0}} \gamma_{ik}
\approx 0
\end{equation}
where we have assumed that both $\tilde{u}^0$ and $\gamma_{ik}$
are approximately constant along the 3-vector $\frac{w^i}{\tilde{u}^0}$,
which lies along the direction of the fluid's rotational velocity 
piece $w^i$. 
Note, that $^{(3)}\!\pounds_{V + \Delta k} w_i$ is of order $O(w)^2$,
while assumptions (\ref{assumption1}) and (\ref{assumption2}) are
$O(1)$ and $O(w)$ in $w^i$. 
Thus alternatively we can view Eq.~(\ref{assumption3})
as an assumption that will hold if $w^i$ is small compared to $D^i\phi$.  
All three assumptions (\ref{assumption1}), (\ref{assumption2})
and (\ref{assumption3}) are discussed in 
appendix \ref{appendix_assumptions} for a simple case.

With the last assumption in Eq.~(\ref{assumption3}) the Euler 
Eq.~(\ref{Euler3}) yields
\begin{equation}
\label{EulerInt}
\frac{h}{u^0} + V^k D_k \phi = -C , 
\end{equation}
where $C$ is a constant of integration, that is in general
different for each star.

In the corotating case where $V^{\mu}=0$, Eq.~(\ref{continuity3}) is
identically satisfied and Eq.~(\ref{EulerInt}) reduces to
\begin{equation}
\label{EulerInt_corot}  
h = -C u^0 ,
\end{equation}
The $u^0$ here can be computed from $u_{\mu}u^{\mu}=-1$ and reduces to
\begin{equation}
u^0 = 1/\sqrt{\alpha^2 - (\beta_i + \xi_i)(\beta^i + \xi^i)} 
\end{equation}
for $V^{\mu}=0$.

If the stars are not corotating $V^{i}$ is given by
\begin{equation}
\label{Vi-eqn}
V^i = \frac{D^i \phi + w^i }{h u^0} - (\beta^i + \xi^i) .   
\end{equation}
In this case the continuity equation~(\ref{continuity3}) becomes
\begin{equation}
\label{continuity4}
D_i \left[ \frac{\rho_0 \alpha}{h}(D^i \phi + w^i) 
          -\rho_0 \alpha u^0 (\beta^i + \xi^i) \right] = 0 . 
\end{equation}
Note that $u_{\mu}u^{\mu}=-1$ yields
\begin{equation}
\label{uzero}
u^0 = \frac{\sqrt{h^2 + (D_i \phi + w_i)(D^i \phi + w^i)}}{\alpha h} ,
\end{equation}
so that Eq.~(\ref{continuity4}) is a non-linear elliptic 
equation for $\phi$. Using $u^0$ from Eq.~(\ref{uzero}) 
the integrated Euler equation~(\ref{EulerInt}) can then be solved for
$h$ with the result
\begin{equation}
\label{h_from_Euler}
h = \sqrt{L^2 - (D_i \phi + w_i)(D^i \phi + w^i)},
\end{equation}
where we use the abbreviations
\begin{equation}
L^2 = \frac{b + \sqrt{b^2 - 4\alpha^4 [(D_i \phi + w_i) w^i]^2}}{2\alpha^2}
\end{equation}
and
\begin{equation}
b = [ (\xi^i+\beta^i)D_i \phi - C]^2 + 2\alpha^2 (D_i \phi + w_i) w^i .
\end{equation}
Note that the rotational piece of the fluid velocity $w^i$ can be freely
chosen, and that
the fluid equations (\ref{continuity4}) and (\ref{h_from_Euler})
reduce to the well known result for 
irrotational stars~\cite{Shibata98,Teukolsky98}
if $w^i=0$.

\subsection{Further simplifications and boundary conditions}
\label{simps_and_BCs}

Next we also choose a maximal slice with $K=0$, and assume that
the conformal 3-metric is flat and given by~\cite{Wilson95,Wilson:1996ty}
\begin{equation}
\label{conflat}
\bar{\gamma}_{ij} = \delta_{ij}.
\end{equation}
This latter assumption merely simplifies our equations
and could in principle be improved by e.g. choosing a 
post-Newtonian expression for $\bar{\gamma}_{ij}$ or by
matching a post-Newtonian metric with a single neutron star solution
similar to~\cite{Tichy02,Yunes:2005nn,Yunes:2006iw,Kelly:2007uc,
JohnsonMcDaniel:2009dq,Kelly:2009js}.
Using Eq.~(\ref{conflat})
the Hamiltonian and momentum constraints in Eq.~(\ref{ham0}) 
and Eq.~(\ref{dK0}) simplify and we obtain
\begin{eqnarray}
\label{ham_mom_dtK0}
\bar{D}^2 \psi &=&
 - \frac{\psi^5}{32\alpha^2} (\bar{L}B)^{ij}(\bar{L}B)_{ij}
 -2\pi \psi^5 \rho \nonumber \\
\bar{D}_j (\bar{L}B)^{ij} &=&
 (\bar{L}B)^{ij} \bar{D}_j \ln(\alpha\psi^{-6})  
 +16\pi\alpha\psi^4 j^i \nonumber \\
\bar{D}^2 (\alpha\psi) &=& \alpha\psi
\left[\frac{7\psi^4}{32\alpha^2}(\bar{L}B)^{ij}(\bar{L}B)_{ij}
      +2\pi\psi^4 (\rho+2S) \right], \nonumber \\
\end{eqnarray}
where
$(\bar{L}B)^{ij} = \bar{D}^i B^j + \bar{D}^j B^i 
- \frac{2}{3} \delta^{ij} \bar{D}_k B^k$,
$\bar{D}_i = \partial_i$, and
\begin{equation}  
B^i = \beta^i + \xi^i + \Omega \epsilon^{ij3} (x^j - x_{CM}^j) .
\end{equation}
The elliptic equations (\ref{ham_mom_dtK0}) have to be solved
subject to the boundary conditions
\begin{equation}
\label{psi_B_alpha_BCs}
\lim_{r\to\infty}\psi = 1, \ \ \ 
\lim_{r\to\infty}B^i = 0, \ \ \
\lim_{r\to\infty}\alpha\psi = 1 
\end{equation}
at spatial infinity.

The equations (\ref{ham_mom_dtK0}) need to be solved together with the fluid
equations (\ref{continuity4}) and (\ref{h_from_Euler}). These fluid
equations simplify in corotating coordinates where $\xi^i=0$. Furthermore
they can be expressed in terms of the derivative operator
$\bar{D}_i$ by noting that
\begin{equation}
D_i \phi = \bar{D}_i \phi , \ \ \
D^i \phi = \psi^{-4} \bar{D}^i \phi .
\end{equation}
In addition $w^i$ can be replaced by
\begin{equation}
w^i = \psi^{-6} \bar{w}^i .
\end{equation}
The latter scaling is useful since
\begin{equation}
D_i w^i = \psi^{-6}\bar{D}_i \bar{w}^i ,
\end{equation}
so that if we choose $\bar{D}_i \bar{w}^i = 0$ we automatically
obtain $D_i w^i = 0$. One obvious choice for the conformal
rotational velocity could be
\begin{equation}
\label{wbar_omega_cross_r}
\bar{w}^i = \epsilon^{ijk} \omega^j (x^k - x_{C*}^k) ,
\end{equation}
where $x_{C*}^k$ is the location of the star center, which could be defined
as the point with the highest rest mass density $\rho_0$ or as the center of
mass of the star. However, it is also possible to choose
\begin{equation}
\bar{w}^i = f(|x^n - x_{C*}^n|) \epsilon^{ijk} \omega^j (x^k - x_{C*}^k) ,
\end{equation}
where $f(|x^n - x_{C*}^n|)$ is any function that only depends on the
conformal distance from the star's center.
Thus the method described here is capable if of giving an arbitrary
rotational velocity to each star.

Also note, that we need a boundary
condition at the star surface to solve Eq.~(\ref{continuity4}). 
This boundary condition can be obtained from Eq.~(\ref{continuity4}) 
itself by evaluating Eq.~(\ref{continuity4}) on the boundary where
$\rho_0 \to 0$ but $\bar{D}_i \rho_0 \neq 0$. Taking this limit
we obtain
\begin{equation}
\label{starBC}
(D^i \phi)D_i \rho_0 + w^i D_i \rho_0 
= h u^0 (\beta^i + \xi^i) D_i \rho_0 
\end{equation}
at the star surface. In applications it may be a good idea to choose
$\bar{w}^i$ such that $\bar{w}^i \bar{D}_i \rho_0$ vanishes, otherwise the
rotational velocity has a component perpendicular to the star's surface.
Also notice that Eq.~(\ref{continuity4}) together with its
boundary condition in Eq.~(\ref{starBC}) do not uniquely specify
the solution. If $\phi$ solves both Eqs.~(\ref{continuity4})
and (\ref{starBC}) $\phi + \mbox{const}$ will be a solution as well.
In numerical codes this kind of ambiguity is usually removed by adding
e.g. the volume integral of $\phi$ over the star to the boundary condition.

\section{The Newtonian Limit}
\label{NewtonLimit}

We now investigate Newtonian limit of the approximate 
matter equations derived above.
If $\varphi$ is the Newtonian potential satisfying 
$\partial_i \partial^i \varphi = 4\pi \rho_0$ 
and $v^i = u^i/u^0$ the Newtonian fluid velocity (in inertial coordinates)
we can express the Newtonian limit as
\begin{eqnarray}
\label{NewtLim}
g_{00} &\to& -1- 2\varphi \nonumber \\
\alpha &\to& 1 + \varphi \nonumber \\
g_{0i} = \beta_i &\to& 0 \nonumber \\
g_{ij} = \gamma_{ij} &\to& \delta_{ij} \nonumber \\
\xi^i &\to& [\Omega\times x]^i \nonumber \\
u^i \to u_i \to \tilde{u}_i \to ^{(3)}\!\tilde{u}_i &\to& v_i \nonumber \\
v_i &\to& \partial_i \phi + w_i \nonumber \\
V^i &\to& \partial^i \phi + w^i - \xi^i \nonumber \\
u^0  &\to&  1 + \frac{v^2}{2} - \varphi \nonumber \\
u_0 = g_{0\mu}u^{\mu} &\to&  -1 - \frac{v^2}{2} - \varphi \nonumber \\
h = 1 + h_N &=& 1 + \epsilon + \frac{P}{\rho_0}
\end{eqnarray}
where $v = \sqrt{v^i v_i}$
and $V^i$ is the fluid velocity in corotating coordinates.

Using Eqs.~(\ref{NewtLim}) Eq.~(\ref{continuity4}) reduces to
\begin{equation}
\partial_i (\rho_0 V^i) = 0
\end{equation}
which is the Newtonian continuity equation in corotating coordinates
where $\partial_{t'} \rho_0 = 0$.

In order to examine the limit of Eq.~(\ref{EulerInt}) we first note that
\begin{equation}
\label{ShibMar1}
\frac{h}{u^0} + ^{(3)}\!\tilde{u}_k V^k = - h u_{\mu}\xi^{\mu} 
\end{equation}
and
\begin{equation}
\label{gradVw}
-D_i(V^k w_k) = V^k (D_k ^{(3)}\!\tilde{u}_i - D_i ^{(3)}\!\tilde{u}_k)
                - ^{(3)}\!\pounds_V w_i .
\end{equation}
Using Eqs.~(\ref{ShibMar1}) and (\ref{gradVw}) together with
the limits in Eqs.~(\ref{NewtLim}) the gradient of
Eq.~(\ref{EulerInt}) yields
\begin{equation}
\label{EulerN1}
\partial_i\left(h_N + \frac{v^2}{2} + \varphi + v_k \xi^k \right)
 + V^k (\partial_k v_i - \partial_i v_k) = ^{(3)}\!\pounds_V w_i .
\end{equation}
In order to show that this is the Euler equation of Newtonian physics
we first note that the time derivative $\partial_{t'}$
in corotating coordinates is related to the 
time derivative $\partial_{t}$ in inertial coordinates 
by $\partial_{t'} =\partial_{t} + ^{(3)}\!\pounds_{\xi}$. Then
\begin{eqnarray}
\partial_{t'} V_i 
&=& \partial_{t'} (\partial_i \phi + w_i-\xi_i)  \nonumber\\
&=& \partial_{t'}(\partial_i \phi)
 + \partial_t w_i + ^{(3)}\!\pounds_{\xi} w_i
 - \partial_{t'} \xi_i \nonumber\\
&=& \partial_{t'}(\partial_i \phi) - \partial_{t'} \xi_i 
 + (\partial_t w_i + ^{(3)}\!\pounds_{\bar{\xi}} w_i) \nonumber\\
&&
 + ^{(3)}\!\pounds_{V+\Delta k} w_i
 - ^{(3)}\!\pounds_V w_i .
\end{eqnarray}
In the last equality all terms but the last vanish,
if we make the same assumptions as in Eqs.~(\ref{assumption1}),
(\ref{assumption2}) and (\ref{assumption3}).
Hence Eq.~(\ref{EulerN1}) can be rewritten as
\begin{equation}
\label{EulerN2} 
\partial_{t'} V_i +  V^k \partial_k V_i + 2[\Omega\times V]_i
+[\Omega\times(\Omega\times x)]_i 
=
-\frac{\partial_i P}{\rho_0} - \partial_i \varphi ,
\end{equation}
which is simply the well known Euler equation of Newtonian physics
expressed in corotating coordinates.
Thus we see that our new matter equations reduce to the correct result
in the Newtonian limit.

\section{Discussion}
\label{discussion}

Realistic neutron stars in binaries will be spinning. From observations 
of millisecond pulsars we know that these spins can be substantial enough
to influence the late inspiral and merger dynamics of the binary.

There have been prior attempts to construct initial data for
spinning neutron stars. In~\cite{Marronetti:2003gk} (hereafter MS) the Euler
equation is not solved directly. Rather it is replaced by an equation
equivalent to 
$ \frac{h}{u^0} + ^{(3)}\!\tilde{u}_k V^k = -C $.
However, as already pointed out by MS, 
this equation agrees with the integrated Euler Eq.~(\ref{EulerInt})
only for the corotating and the irrotational case. Thus in general the Euler
equation is violated in the MS approach.
Furthermore, MS split $u^i/u^0$ and not $^{(3)}\!\tilde{u}^i$
into an irrotational and a rotational part (see Eq.~(\ref{utilde-split})). 
This has two consequences. First, their equations do not have the
correct limit in the irrotational case. And second, since $u^{\mu}/u^0$
is not a purely spatial vector it is inconsistent to set $u^i/u^0$ 
equal to something like $D^i \phi$ which is a purely spatial vector.
This explains why the continuity equation of MS has no shift terms unlike
in Eq.~(\ref{continuity4}) and in~\cite{Shibata98,Teukolsky98}.
When MS compare their results for a particular corotating case
with~\cite{Uryu:1999uu} they find that their approach introduces
errors of about 2\% in the angular momentum.

Another approach to include spin that is aligned with the orbital angular
momentum was proposed in~\cite{Baumgarte:2009fw}, hereafter BS.  This
approach does not seek to analytically integrate the Euler equation as we
have done here.  Instead the divergence of Eq.~(\ref{Euler2}) is set to
zero, which leads to another elliptic equation.  However, as pointed out
first by Gourgoulhon~\cite{Baumgarte:2009fwErr}, 
in general the Euler equation itself is not satisfied
if we only enforce its divergence to be zero.  Hence the BS approach can
lead to initial data that do not obey the Euler equation.  Furthermore, the
boundary condition given by BS for their new elliptic equation seems to
imply that the star surface is always at the same location. 
If we consider the usual numerical treatment where we start from an
initial guess for the stars which is iteratively refined, it is unclear
how the star surface can change during the iterations.

The purpose of this paper is thus to introduce a new method
for the computation of binary neutron star initial data with
arbitrary rotation states. Our method is derived from the standard
matter equations of perfect fluids together with certain quasi-equilibrium
assumptions. We assume that there is an approximate helical Killing vector
$\xi^{\mu}$ and that Lie derivatives of the metric variables with respect to 
$\xi^{\mu}$ vanish. We also assume that scalar matter variables 
such as $h$ or $\rho_0$ have Lie derivatives that vanish with respect to 
$\xi^{\mu}$. However, as discussed in appendix \ref{appendix_assumptions}
the Lie derivative of the fluid velocity $u^{\mu}$ is expected to be
non-zero for arbitrary spins. We split the fluid velocity
$u^{\mu}$ into an irrotational piece (derived from a potential $\phi$)
and a rotational piece $w^i$, and assume that only
the irrotational piece has a vanishing Lie derivative (see
Eq.~(\ref{assumption1})) with respect to $\xi^{\mu}$. This can be
interpreted the natural generalization of the irrotational case where one
commonly assumes $\pounds_{\xi} h u^{\mu} = 0 $. 
Furthermore we know that the spin of each star remains approximately
constant since the viscosity of the stars is insufficient for tidal
coupling~\cite{Bildsten92}. To incorporate this fact,
we use Eq.~(\ref{assumption2}) which is based on the assumption
that $w^i$ is constant along the star's motion described by the irrotational
velocity piece $\nabla^{\mu} \phi$. Since $\nabla^{\mu} \phi$ is equivalent
to the velocity of the star center, this latter assumption captures the fact
that the spin or rotational velocity $w^i$ of each star remains approximately
constant. With these two assumptions the Euler equation simplifies to
Eq.~(\ref{Euler3}). 
In order to analytically integrate Eq.~(\ref{Euler3}) we use the additional
assumption (\ref{assumption3}) that $\tilde{u}^0$ and $\gamma_{ij}$ are
constant along the field lines of the rotational velocity piece.
We then arrive at the two matter equations
Eq.~(\ref{continuity4}) and (\ref{h_from_Euler}). These equations reduce
to well known equations~\cite{Shibata98,Teukolsky98} 
for the irrotational case of $w_i=0$. They also reduce to the
corotating limit (where $V^i=0$) as is evident from Eqs.~(\ref{continuity3})
and (\ref{EulerInt}) which are written in terms of $V^i$.
Furthermore, our equations reduce to the correct Newtonian limit.

The elliptic equation in Eq.~(\ref{continuity4}) can to solved (for $\phi$)
together with the Eqs.~(\ref{ham_mom_dtK0}) for the metric variables once
the enthalpy $h$ is known.  However, the enthalpy given by
Eq.~(\ref{h_from_Euler}) depends on the metric variables, $\phi$ and $w^i$.
Apart from their dependence on $w^i$ this set of equations has a similar
structure as for the case of irrotational neutron stars (where
$w^i$ vanishes). The standard way (see e.g. ~\cite{Tichy:2009yr}) to solve
such a mixture of elliptic and algebraic equations is by iteration, where
at each step we first solve the elliptic equations for a given $h$
and then use the algebraic Eq.~(\ref{h_from_Euler}) to update $h$.
At each step we also need to specify $w^i$. One way to do this would be
by choosing a constant $\bar{w}^i$ as in Eq.~(\ref{wbar_omega_cross_r}).
Note however, that other choices for $w^i$ are possible.
We plan to investigate these possibilities in future numerical studies
of our new method. For such studies it might useful to use a numerical code
like LORENE~\cite{Grandclement-etal-2000:multi-domain-spectral-method,
Gourgoulhon:2000nn,Grandclement02,lorene_web} 
or SGRID~\cite{Tichy:2006qn,Tichy:2009yr,Tichy:2009zr}, where the star
surface is always at a domain boundary so that the boundary
condition in Eq.~(\ref{starBC}) can be easily implemented.

\begin{acknowledgments}

It is a pleasure to thank Pedro Marronetti
for helpful discussions.
This work was supported by NSF grant PHY-0855315.


\end{acknowledgments}

\appendix

\section{The matter quasi-equilibrium assumptions in a simplified case}
\label{appendix_assumptions}

In the Newtonian limit $^{(3)}\!\tilde{u}_i$ is the equal to the fluid
velocity in the inertial frame.
When the two stars are well separated it is clear that 
each star is well approximated by an orbiting and spinning sphere.
In this case the fluid velocity inside a star is given by
\begin{equation}
^{(3)}\!\tilde{u}_i 
\approx [\Omega \times x_{C*}]_i + [\omega \times (x-x_{C*})]_i ,
\end{equation}
where $x_{C*i}$ is the (time dependent) 
location of the star center, and $\Omega_i$ and $\omega_i$ are
the angular velocities of the orbital and spinning motion
of the star. Within this approximation we then have
\begin{equation}
\phi \approx  [\Omega \times x_{C*}]_k x^k, \ \ \ \
w_i  \approx  [\omega \times (x-x_{C*})]_i .
\end{equation}
It is then easy to verify that the assumptions in 
Eqs.~(\ref{assumption1}) and (\ref{assumption2}) are identically satisfied.
Furthermore for approximate spherical symmetry
we see that the assumptions in Eq.~(\ref{assumption3}) hold as well.
In addition, we find that
\begin{equation}
\Delta k_i = [\Omega \times x]_i - D_i \phi = [\Omega \times (x-x_{C*})]_i .
\end{equation}
From this it follows that
\begin{equation}
\gamma_i^{\nu}\pounds_{\xi} \tilde{u}_{\nu} 
= ^{(3)}\!\pounds_{\Delta k} \tilde{w}_{i} 
= (\Omega_i \omega_j - \omega_i \Omega_j) (x^j-x_{C*}^j)
\end{equation}
which illustrates that $\pounds_{\xi} \tilde{u}_{\nu}$ does not vanish even 
in this simplified case. The only case when 
$\pounds_{\xi} \tilde{u}_{\nu}$ can vanish is if the spin is aligned with
the orbital angular momentum, i.e. if $\omega_i = a \Omega_i$ for some 
constant $a$.


\bibliography{references}

\end{document}